# Photostable molecules on chip: integrated single photon sources for quantum technologies

P. Lombardi,[†,‡,⊥] A. P. Ovvyan,[¶,‡,⊥] S. Pazzagli,[§,†] G. Mazzamuto,[†,‡] G. Kewes,[k] O. Neitzke,[k] N. Gruhler,[¶] O. Benson,[k] W.H.P. Pernice,[¶] F. S. Cataliotti,[§,‡,†] and C. Toninelli[*,†,‡]

†CNR-INO, Istituto Nazionale di Ottica, Via Carrara 1, 50019 Sesto F.no, Firenze, Italy
‡LENS, Via Carrara 1, 50019 Sesto F.no, Firenze, Italy
¶Institute of Physics, University of Muenster, Muenster, Germany
§Università di Firenze, Via Sansone 1, I-50019 Sesto F.no, Firenze, Italy
kNano-Optik, Institut für Physik, Humboldt-Universität zu Berlin, Berlin, Germany
⊥Contributed equally to this work

E-mail: toninelli@lens.unifi.it
Phone: +39 055-4572134

## Abstract

The on-chip integration of quantum light sources and nonlinear elements poses a serious challenge to the development of a scalable photonic platform for quantum information and communication. In this work we demonstrate the potential of a novel hybrid technology which combines single organic molecules as quantum emitters and dielectric chips, consisting of ridge waveguides and grating far-field couplersDibenzoterrylene molecules in thin anthracene crystals exhibit long-term photostability, easy fabrication methods, almost unitary quantum yield and life-time limited emission at cryogenic temperatures. We couple such single emitters to silicon nitride ridge waveguide with a coupling efficiency of up to 42 ± 2 %, considering both propagation directions. The platform is devised to support both on-chip and free-space single photon processing.

Optical quantum technologies rely on efficient sources of non-classical light.[1,2] In particular, bright and reliable sources of single photons are a key ingredient for linear optical quantum computing (LOQC),[3] boson sampling algorithms, secure quantum communications[4] and quantum imaging.[5] In this context, single quantum emitters (QEs) under pulsed excitation have been proposed as deterministic sources of indistinguishable single photons[6] They represent an alternative route to heralded photon generation from parametric down conversion,[7] insofar as decoherence and losses are suppressed. Enhancement and control of light-matter interaction through dedicated interfaces is hence crucial, allowing for efficient collection and processing of the delivered photons. Optical cavities for example are employed to improve brightness and indistinguishability of single photon sources,[8–10] allowing as well for optical non-linearities in the few photon regime.[11,12] The nanoscale field confinement in dielectric and plasmonic antennas has also been put forward[13] to yield unidirectional emission,[14,15] almost unitary collection efficiency[16] and fast repetition rates.[17] On the other hand, long-distance quantum communication and LOQC require the coupling of single photons to well-defined propagating modes, i.e. to optical fibers[18,19] or in integrated photonic platforms.[20] In particular, direct coupling of QEs to single mode waveguides (WGs) in a planar geometry is favorable for realizing integrated quantum photonic circuits,[21–24] targeting also logic operations[25,26] and single photon detection on the same chip. In this respect, high coupling efficiency can be obtained in photonic-crystal WGs,[27,28] whereas on-chip ridge geometries[29–31] allow for long-range propagation and direct signal processing. Notably, positioning QEs in the pronounced evanescent field attainable in ridge WGs has been proposed for efficient collection into on-chip guided modes.[29,32?]

We demonstrate here the emission of single photons from Dibenzoterrylene (DBT) molecules into $Si_3N_4$ ridge WGs at room temperature. DBT molecules embedded in a rigid matrix of crystalline anthracene (Ac) show photostable emission both at room and at cryogenic temperatures. Moreover, at 3 K, the emitted photons can be made indistinguishable with 30 % emission into the zero phonon line, and are therefore well suited for carrying quantum information with high fidelity.[33] The DBT:Ac system holds promise because of an almost unitary quantum yield and negligible blinking, limited by the intersystem crossing yield ($10^{-7}$) to the short-lived triplet state.[34,35] This behavior is preserved also in crystals as thin as 50 nm, obtained by drop-casting DBT:Ac in solution onto a rapidly spinning plate, thus enabling a straight-forward integration[36] (see also SI). In this work, nanoscale emitters are spin-coated onto the photonic chip, containing WGs and far-field couplers as sketched in Fig.1a. DBT single photon emission is hence coupled to the evanescent tail of the WG guided mode, with a molecule-to-WG coupling efficiency ($\beta$) of up to $42 \pm 2\%$. The overall single-photon source efficiency, including emission into the guided mode, propagation with $4.9-dB/cm$ losses,[37] and emission into a quasi-gaussian mode in free space is estimated to be around $16\%$. These results are competitive with state-of-the-art single photon emission into ridge WGs' modes from other solid state systems,[31,38,39] while offering a novel platform with high versatility. The advantages of our approach include a small footprint, simple fabrication methods and scalability towards arrays of integrated single photon sources.

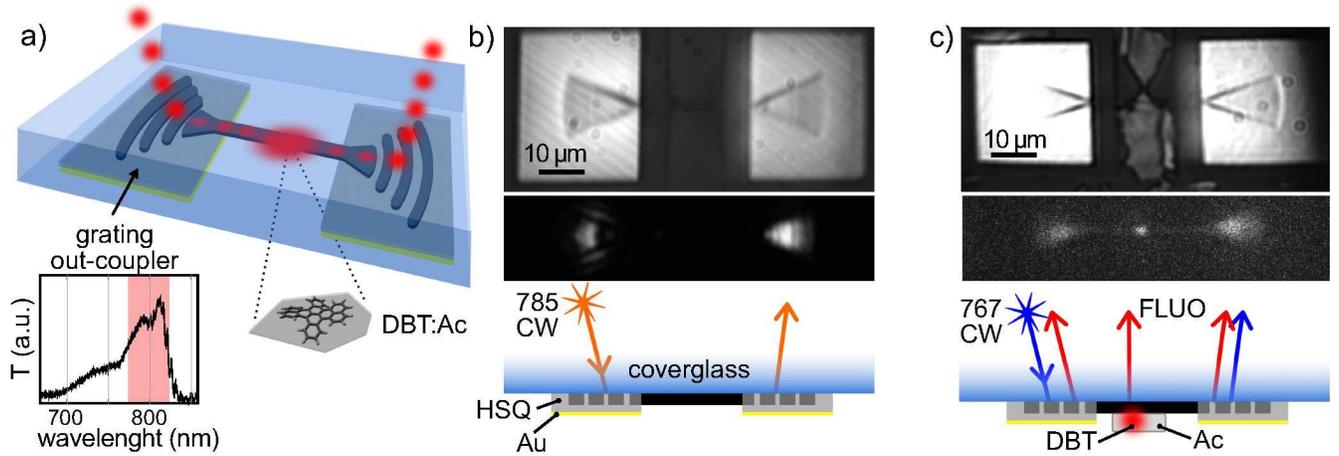

Figure 1: **Single-molecule-device working principle.** a) Cartoon of the device showing a $Si_3N_4$ waveguide with two focussing grating outcouplers, coated with an HSQ layer and gold. A single molecule embedded in a thin anthracene crystal laying on top of the bare waveguide region emits a train of guided single photons. In the inset, the grating optical response shows broadband operation around the DBT fluorescence window (red shaded area). In panel b) and c) from bottom to top we report a sketch of the excitation configuration, the resulting signal onto the EMCCD camera and white light images for an empty device (b) and a hybrid molecule-on-waveguide system (c).

The substrate for fabricating our optical chips consists of a square glass die, covered with a $175-nm$ thin layer of silicon nitride ($Si_3N_4$). Nanophotonic circuits are then fabricated using three steps of electron-beam lithography, followed by reactive ion etching, yielding $500\,nm$-wide single mode optical WGs. In order to couple light in and out of the WGs, we employ focusing grating couplers as described in Ref.[40] The gratings are fully-etched, buried in a $760\,nm$-thick hydrogen silsequioxane (HSQ) buffer layer with a $120\,nm$-thick gold mirror on top to enhance directionality. For this design, finite elements simulations yield a value for the output coupling efficiency ($\eta_c$) of up to $90\%$, (see SI). A typical white light image of the bare waveguide-coupler device is shown in the top panel of Fig.1b, where the gold mirrors mark the coupler regions. In our design, apodization and tapering length have been optimized to support a gaussian mode in free space with a full width at half maximum ($FWHM$) $\simeq 4\,\mu m$ (see SI, fig.4). We hence extract the outcoupling efficiency from transmission measurements, as depicted in the bottom panel of Fig.1b. The transmission of a laser at $785\,nm$ is collected by the imaging system (Fig.1b middle panel), integrated over the coupler area and normalized to the reflection from a silver mirror (after background subtraction on both images). The single coupler efficiency is then evaluated as the square root of the throughput. Measured values yield on average $\eta_c = 35 \pm 5\%$. We attribute the discrepancy with the simulation results to the non-optimal HSQ thickness in the fabricated chips. The transmission spectrum in Fig.1a further shows the coupler bandwidth of about $50-nm$. DBT-doped anthracene crystals are deposited on the chip by drop casting. The results of this procedure are then examined by white light images, such as the one in the top panel of Fig.1c. In this case for example just a tiny portion of the crystal covers the waveguide central region. As a first experiment, the pump laser at 767 nm is coupled to the WG through the left grating coupler. Molecules in the vicinity of the WG are then excited by the evanescent tail of the guided pump light. The emitted fluorescence partially couples back into the WG and propagates in two opposite directions towards the grating couplers (see the cartoon in Fig. 1c). The whole device is imaged onto an electron multiplying charge coupled device (EM-CCD), after filtering out the pump light. As a first evidence of DBT-WG coupling, we show in the middle panel of Fig.1c clear signature of the expected fluorescence from the grating couplers.

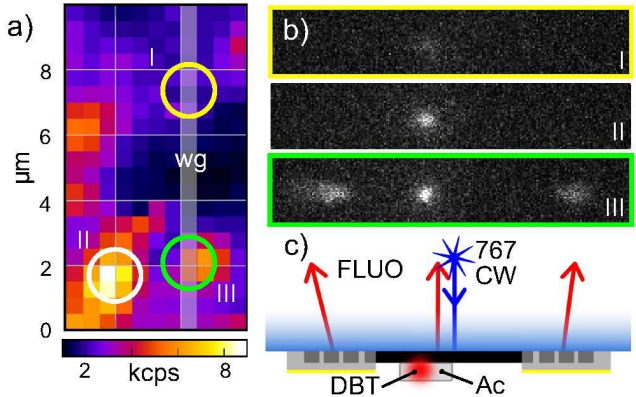

Figure 2: **Guided fluorescence.** a) Confocal fluorescence map from DBT:Ac crystals, deposited on the chip around a waveguide region. b) EMCCD image of the fluorescence from the entire device when excited with a diffraction-limited laser spot in point I (no molecule on WG), II (molecule off guide) and III (WG-coupled molecule). c) Cartoon for the experimental configuration.

In the second round of experiments, in order to better isolate the molecular emission, pump and fluorescence are spatially separated. We first determine the spatial distribution of molecules within the anthracene crystal by scanning the sample under the diffraction-limited laser spot and detecting fluorescence from the same position with a single photon avalanche diode (SPAD). An example of such a fluorescence map is displayed in Fig.2a, where the grey-shaded rectangle indicates the underlying waveguide structure. In this case, several molecular emitters can be identified within the scanned region. We then excite with the laser spot at three different fixed positions and record the corresponding fluorescence images from the whole device surface with the EMCCD camera (panels in Fig.2b). Panel II and III correspond to the excitation of a molecule sitting far and in the near-field of the waveguide, respectively. Clearly, the output couplers light up when a molecule is optically coupled to the waveguide, whereas direct scattering between the confocal fluorescence spot and the gratings is negligible. Panel (I) shows the background signal, which is obtained by illuminating a crystal without DBT molecules on the WG. Some residual light is present on the illumination point and vanishes at the couplers. This can be associated with some residual fluorescence from the anthracene matrix and is about a factor 5 smaller than the typical signal from a single DBT molecule. The relative intensity within panel (III) is determined by the probability of emission into the guided mode, the efficiency of the grating couplers and the collection efficiency of the objective. These parameters are estimated further in the text. We note that the fluorescence signal in Fig.2a is stronger for molecule II than for molecule III, in contrast to what displayed in Fig.2b. The asymmetry is due to the different polarization of the pump light, adjusted first to match the field in the guided mode and then optimized for maximum signal in the EMCCD at the coupler position, allowing for an extra-tilt usually smaller than 20°.

We then analyze the quantum nature of the WG-coupled light by measuring the field second order autocorrelation function ($g^{(2)}(\tau)$), presented in Fig.3c and d. Here, coincidence counts are plotted for fluorescence photons collected either from the source position or from one coupler. The dashed circles in Fig.3a approximately indicate the corresponding selected areas in detection, obtained by shifting the position and changing the size of a spatial filter. The fluorescence map for the same device is further shown in Fig.3b. The experimental data for the $g^{(2)}(\tau)$ (black solid line) are presented without background subtraction and show good agreement with the fitting function $g^{(2)}(\tau) = 1 - b\exp(\tau/T)$ (red solid line). Clear evidence of single photon emission is obtained in both cases. Indeed, extracting $g^{(2)}(0)$ as a free fitting parameter with the relative standard error, we get $g^{(2)}(0) = 0.33 \pm 0.09$ when collecting directly from the source and $g^{(2)}(0) = 0.50 \pm 0.05$ for light scattered by a single coupler. The somewhat higher value of $g^{(2)}(0)$ at the coupler is attributed to the lower signal to noise ratio determined also by the larger size of the collection region.

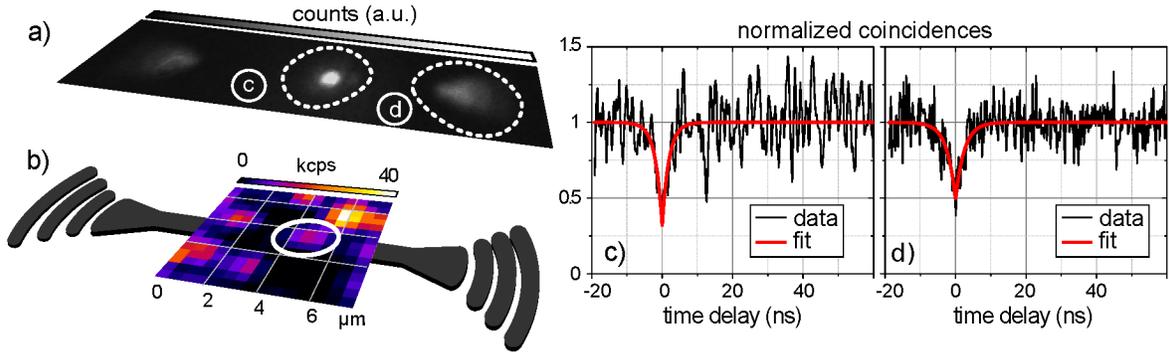

Figure 3: **Integrated single photon source.** In panel a) a fluorescence confocal scan allows to isolate possibly coupled molecules. The highlighted molecule fluorescence imaged onto the EMCCD camera in b) shows signal at the outcouplers. The coupling of single photons into the waveguide finds evidence in the antibunching dip relative to the confocal spot c) and to the coupler area d), respectively.

In table 1 we summarize the main results obtained on the illustrative device relative to Fig.3, providing WG-coupled single photon emission. Processing the direct measurements of the detected photon count rate integrated over the two couplers ($S_c$) and the $g^2(0)$ by using the data in the table, we estimate the most relevant parameters for this device, highlighted in grey. In particular the on-chip source purity ($g^{(2)}_{\text{on}}(0)$) is crucial, as operations with single photons are ideally performed on the platform before off-chip read-out. Therefore we need to evaluate and eliminate the contribution of the background signal (B), given by the residual light which is not guided. To do so, we excite a spot far from any WG and molecule, and collect light with the SPAD from a reference area, which is at the same relative distance from the laser spot and of the same size, as the one selected for measuring the $g^2$ function at the coupler. We can then calculate the probability of obtaining a click in the detector from guided light as $p = (S_c - B)/S_c = 79\%$. The on-chip source purity is then given by $g^{(2)}_{\text{on}}(0) = 1 + \frac{g^{(2)}(0)-1}{p^2} = 0.02 \pm 0.12$, representing an extremely low multi-photon probability inside the WG.

A further key figure of merit for integrated single photon sources is the emitter-WG coupling efficiency ($\beta$ factor), defined as $\beta = \frac{\Gamma_{\text{wg}}}{\Gamma_{\text{wg}}+\Gamma_{\text{free}}+\Gamma_{\text{nr}}}$, where $\Gamma_{\text{wg}}$ and $\Gamma_{\text{free}}$ are the radiative decay rate into the waveguide mode and in all other optical modes, respectively, while $\Gamma_{\text{nr}}$ accounts for the non-radiative decay channels. We can obtain the value of $\beta$ in two independent ways. First, $\beta_{\text{est}}$ is determined by comparing the measured count rate from both out-couplers ($48 \pm 4\,\text{kcps}$) with the estimated emission rate from the molecule ($\frac{1}{2\tau}\frac{s}{s+1}$, where $\tau$ is the excited state lifetime and $s$ the saturation parameter), taking into account all other losses in the system (see formula in Table 1). For the fluorescence lifetime of WG-coupled molecules we find an average value of $4.2 \pm 0.4\,\text{ns}$, which is comparable to the case of uncoupled ones[35] within the experimental uncertainty of $10\,\%$. This suggests also that the interaction with the WG only weakly enhances the emission rate, as observed in Ref.[29] as well.

We experimentally estimate $\eta_c = 25 \pm 2\,\%$

Table 1: Performances and efficiencies of the same device as in Fig.3.

| | | | |
|---|---|---|---|
| $g^2(0)$ | off-chip purity | measured | 0.33 ± 0.09 |
| $S_c$ | fluo. signal | measured | 48 ± 4 kHz |
| B | background | measured | 10 ± 2 kHz |
| τ | Lifetime | measured | 4.2 ± 0.4 ns |
| QY | quantum yield | Ref [41] | ≃ 95 % |
| s | saturation | estimated | ≃ 1/5 |
| $\eta_c$ | coupler eff. | measured | 25 ± 2 % |
| $\eta_{opt}$ | optics trans. | measured | ≃ 10 % |
| $\eta_{det}$ | detector eff. | specs | ≃ 50 % |
| $\eta_f$ | coll. efficiency from exc. spot | simulations | 5.0 ± 1.5 % |
| $g^2_{on}(0)$ | on-chip purity | $1 + \frac{g^{(2)}(0)-1}{[(S_c-B)/S_c]^2}$ | 0.02 ± 0.12 |
| $\beta_{est}$ | SM-WG coup. | $\frac{(S_c-B)\,2\tau\,(s+1)/s}{QY\,\eta_c\,\eta_{opt}\,\eta_{det}}$ | ≃ 17 % |
| $\beta_{meas}$ | SM-WG coup. | $\frac{\check{S}_c/\eta_c}{\check{S}_c/\eta_c + \check{S}_f/\eta_f}$ | 20 ± 5 % |

for the specific device (with uncertainty due to inaccuracy in the laser in-coupling and evaluated as a one standard deviation value) and the transmission through filters+optics to be about 10 %. Assuming then a quantum yield of approximately 95 %, a saturation parameter s ≃ 1/5 and a detector efficiency $\eta_{det}$ = 50 %, we obtain $\beta_{est}$ ≃ 17 %. Alternatively, the beta-factor can also be determined by comparing the fluorescence intensity on the EM-CCD camera in the coupler areas $\check{S}_c$ with the molecule residual emission into free space $\check{S}_f$, taking into account the corresponding collection efficiencies (η)[a]. With these definitions, the coupling probability into the WG can be obtained as $\beta_{meas} = \frac{\check{S}_c/\eta_c}{\check{S}_c/\eta_c + \check{S}_f/\eta_f}$. For light coming from the WG-guided mode, the collection efficiency is determined by the efficiency of the grating coupler only, since the scattered mode entirely fits within the objective numerical aperture (NA). In contrast, the collection efficiency $\eta_f$ for the residual light emitted into free space is dependent on the radiation pattern of the emitter and the NA of the objective. For the estimation of this geometrical parameter, we rely on numerical simulations (see SI), confirmed also by semi-analytical calculations for the radiation pattern of a Hertzian dipole embedded in the 2D multilayer system which represents the sample.[15] We find $\eta_f$ = 5.0 ± 1.5 %, where the error bar is estimated by varying the dipole position and orientation within the anthracene layer. Using such estimate for $\eta_f$, we then determine $\beta_{meas}$ solely from EMCCD images, obtaining a value of 20 ± 5 % for the device discussed in Fig.3 and in Table 1. We can hence conclude that the two determination of the molecule-WG coupling efficiency are consistent. We note also that the latter method is more trustworthy, being independent from the setup response function and from the molecule QY.

Using the second approach we then characterized several devices, obtaining a range of different coupling efficiencies with a peak value of $\beta_{meas}$ = 42 ± 2 %. Further analysis on the spatial dependence of the coupling mechanism for the device with the best performances is provided in the SI. The variability in $\beta_{meas}$ can be attributed to the possible span in the crystal thickness, to the random in-plane orientation of the molecule with respect to the WG mode and also to the different molecule position within the anthracene crystal. In order to further quantify the origin of this variation, we performed additional FEM simulations in 3D. The simulated structure is outlined in Fig.4a (see SI for details): the anthracene layer thickness is varied within the experimentally estimated range. We assume for the refractive indices $n_{Ac}$ = 1.8, $n_{Si_3N_4}$ = 2 and $n_{glass}$ = 1.51. The DBT molecule is modeled as a Hertzian dipole, placed in the anthracene layer at a variable vertical distance from the top surface ($d_y$) and horizontally at the center of the waveguide, with orientation $E_x$ parallel to the most prominent field component of the only guided mode. In panels b, c, and d we report the electric field norm for a crystal thickness (h) of 100 nm and $d_y$ = 10 nm in three sections of the simulation domain, as marked in the cartoon. From these simulations the β factor can be extracted as the ratio between the Poyinting vector flux through the WG (the detector area being the one depicted in Fig.4d) and the overall radiated power. In Fig.4e we plot the such-calculated $\beta$ as a function of the dipole distance to the waveguide top surface (black dots) and, for a given $d_y$, as a function of the anthracene crystal thickness (colored dots). We observe that the experimental results are correctly described by the numerical model, when the variation in the coupling efficiency is accounted for by different possible values of $d_y$ and $h$. The total radiated power calculated from the simulations is also consistent with the lifetime measurements, resulting in values only 10 % higher when the molecule is in close proximity to the WG. This suggests that the attained $\beta$ factor should be interpreted as a redistribution of the emission into the guided mode of the field otherwise radiated at high polar angles in the glass substrate.

---

[a]Background subtraction is performed against an average of reference frames, corresponding to positions on the crystal without molecules.

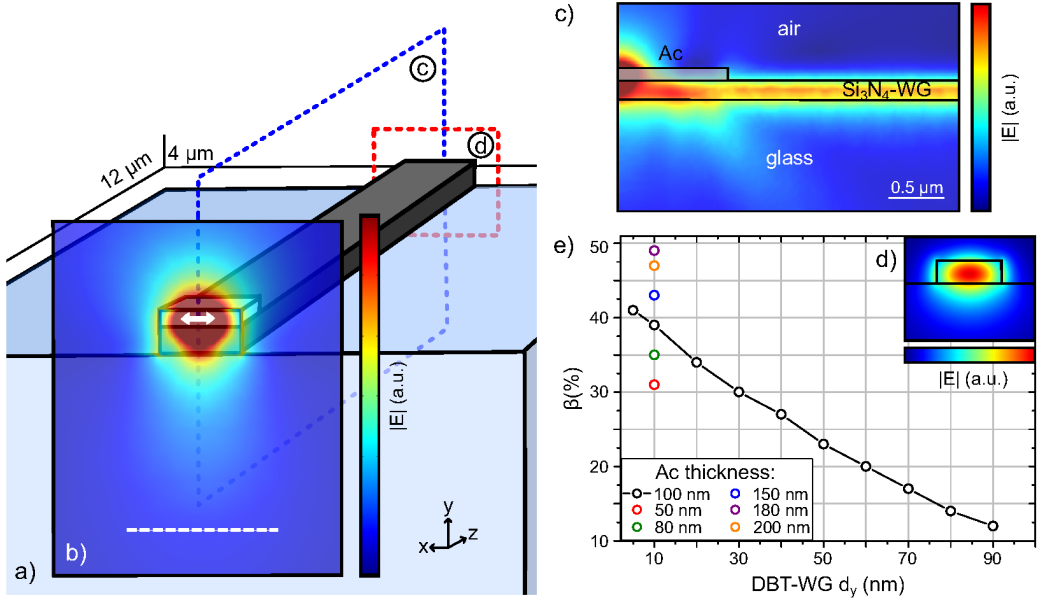

Figure 4: **Numerical simulations.** a) Layout of the 3D numerical simulations displaying half of the waveguide total length (24 µm). A dipolar emitter (white arrow) mimicking a single molecule is placed in a 100 nm-thick anthracene crystal (n=1.8). b) Plot of the electric field norm on the plane containing the dipole. The dashed white line represents the objective acceptance angle. Panels c) and d) display the electric field norm in the two respective waveguide sections. In e) the coupling efficiency into the waveguide mode is reported as a function of the dipole distance to the WG top surface (black dots) and for different crystal thicknesses (colored dots).

In conclusion we have observed single molecule emission into an integrated optical circuit, consisting of a ridge WG terminated with two grating far-field couplers. The best measured performances and efficiencies are highlighted in Table 2, where we also extrapolate values at saturation for the on-chip single photon rate ($S_{\text{on}}^{\text{sat}}$) and off-chip source efficiency (BR$_{\text{off}}^{\text{sat}}$). $S_{\text{on}}^{\text{sat}}$ is given by $\frac{1}{4\tau} QY \beta_{\text{meas}}$ and amounts to approximatively 24 MHz. We note that this value corresponds to emission into the two WG opposite directions and hence -in principle- to a source of single-photon entangled states.[42] Besides on-chip operation, the same device competitively performs also as single photon source into the free-space quasi-gaussian mode of the grating coupler. The off-chip source brightness can then be defined as the probability upon excitation to emit a photon into the guided mode, which is then scattered by the grating. Estimated as BR$_{\text{off}}^{\text{sat}} = QY \beta_{\text{meas}} \eta_c$, we obtain a value around 16 %. The simplicity of the fabrication process combined with the unique optical properties of the emitter at low temperatures make our system a good source candidate for scalable approaches to on-chip quantum computation. Moreover, a straight-forward extension to many

Table 2: Best measured efficiencies and extrapolated performances at saturation for our integrated single photon source, based on a single DBT:Ac molecule coupled to a ridge SiN WG.

| | | | |
|---|---|---|---|
| $\eta_c$ | coupler efficiency | measured | 40 ± 2 % |
| $\beta_{meas}$ | SM-WG coupling | $\frac{\check{S}_c/\eta_c}{\check{S}_c/\eta_c + \check{S}_f/\eta_f}$ | 42 ± 2 % |
| $S_{on}^{sat}$ | photon flux on-chip @ sat. | $\frac{1}{4\tau}$ QY $\beta_{meas}$ | ≃ 24 MHz |
| $BR_{off}^{sat}$ | off-chip brightness | QY $\beta_{meas}$ $\eta_c$ | ≃ 16 % |

QEs in a single one-dimensional channel could allow the study of manybody effects and quantum correlations.[27,43,44] In order to enhance the coupling efficiency to the WG mode while keeping on-chip losses small, more complex photonic designs can be envisioned, which are compatible with the same fabrication procedure. Replacing one output coupler with a Bragg mirror at the appropriate distance would readily provide a factor of 4 improvement,[29] whereas inscribing a slot into the ridge WG would yield a higher field concentration. Finally, we are considering breaking the translational invariance and engineering a higher group index for the same mode area with corrugated-type waveguides,[45] or hybrid photonic crystal ones, yielding high effective photon masses.[46]

Acknowledgement The authors would like to thank F. Sgrignuoli for helping with numerical simulations, B. Tiribilli, for inspection of the samples by atomic force microscopy, D.S. Wiersma for access to clean room facilities, M. Bellini and C. Corsi for Ti:sapphire operation and Silvia Diewald for help with electron beam lithography. This work benefited from the COST Action MP1403 (Nanoscale Quantum Optics). It is supported by the Erasmus Mundus Doctorate Program Europhotonics (Grant No. 159224-1-2009-1-FR-ERA MUNDUS-EMJD), by the Fondazione Cassa di Risparmio di Firenze (GRANCASSA), MIUR program Atom-Based Nanotechnology and by the Deutsche Forschungs Gemeinschaft (DFG) through the sub-projects B2 and B10 within the Collaborative Research Center (CRC) 951 (HIOS).

The authors declare no competing financial interest.

Supporting Information. Details about the optical setup, the sample characterization, the fabrication method and the numerical simulations.

## References

(1) O'Brien, J. L.; Furusawa, A.; Vuckovic, J. Photonic quantum technologies. Nat. Phot. 2009, 3, 687–695.

(2) Walmsley, I. A. Quantum optics: Science and technology in a new light. Science 2015, 348, 525–530.

(3) Knill, E.; Laflamme, R.; Milburn, G. J. A scheme for efficient quantum computation with linear optics. Nature 2001, 409, 46–52.

(4) Azuma, K.; Tamaki, K.; Lo, H.-K. All-photonic quantum repeaters. Nature Communications 2015, 6, 6787–.

(5) Gatto Monticone, D.; Katamadze, K.; Traina, P.; Moreva, E.; Forneris, J.; Ruo-Berchera, I.; Olivero, P.; Degiovanni, I. P.; Brida, G.; Genovese, M. Beating the Abbe Diffraction Limit in Confocal Microscopy via Nonclassical Photon Statistics. Phys. Rev. Lett. 2014, 113, 143602.

(6) Lounis, B.; Moerner, W. E. Single photons on demand from a single molecule at room temperature. Nature 2000, 407, 491–493.

(7) Mosley, P. J.; Lundeen, J. S.; Smith, B. J.; Wasylczyk, P.; U'Ren, A. B.; Silberhorn, C.; Walmsley, I. A. Heralded Generation of Ultrafast Single Photons in Pure Quantum States. Phys. Rev. Lett. 2008, 100, 133601.

(8) Somaschi, N. et al. Near-optimal single-photon sources in the solid state. Nat Photon 2016, 10, 340–345.

(9) Faraon, A.; Barclay, P. E.; Santori, C.; Fu, K.-M. C.; Beausoleil, R. G. Resonant

enhancement of the zero-phonon emission from a colour centre in a diamond cavity. Nat Photon 2011, 5, 301–305.

(10) Riedrich-Möller, J.; Arend, C.; Pauly, C.; Mücklich, F.; Fischer, M.; Gsell, S.; Schreck, M.; Becher, C. Deterministic Coupling of a Single Silicon-Vacancy Color Center to a Photonic Crystal Cavity in Diamond. Nano Lett. 2014, 14, 5281–5287.

(11) Tiecke, T. G.; Thompson, J. D.; de Leon, N. P.; Liu, L. R.; Vuletic, V.; Lukin, M. D. Nanophotonic quantum phase switch with a single atom. Nature 2014, 508, 241–244.

(12) Giesz, V.; Somaschi, N.; Hornecker, G.; Grange, T.; Reznychenko, B.; De Santis, L.; Demory, J.; Gomez, C.; Sagnes, I.; Lema?ï£¡tre, A.; Krebs, O.; Lanzillotti-Kimura, N. D.; Lanco, L.; Auffeves, A.; Senellart, P. Coherent manipulation of a solid-state artificial atom with few photons. Nature Communications 2016, 7, 11986–.

(13) Chang, D. E.; Sorensen, A. S.; Demler, E. A.; Lukin, M. D. A single-photon transistor using nanoscale surface plasmons. Nat Phys 2007, 3, 807–812.

(14) Curto, A. G.; Volpe, G.; Taminiau, T. H.; Kreuzer, M. P.; Quidant, R.; van Hulst, N. F. Unidirectional Emission of a Quantum Dot Coupled to a Nanoantenna. Science 2010, 329, 930–933.

(15) Checcucci, S.; Lombardi, P.; Rizvi, S.; Sgrignuoli, F.; Gruhler, N.; Dieleman, F. B. C.; Cataliotti, F. S.; Pernice, W. H. P.; Agio, M.; Toninelli, C. Planar optical antenna to direct light emission. Light: Science and Applications 2016, 6, e16245.

(16) Chu, X.-L.; Brenner, T. J. K.; Chen, X.-W.; Ghosh, Y.; Hollingsworth, J. A.; Sandoghdar, V.; Goetzinger, S. Experimental realization of an optical antenna designed for collecting 99% of photons from a quantum emitter. Optica 2014, 1, 203–208.

(17) Hoang, T. B.; Akselrod, G. M.; Mikkelsen, M. H. Ultrafast Room-Temperature Single Photon Emission from Quantum Dots Coupled to Plasmonic Nanocavities. Nano Lett. 2015, 16(1), 270–275.

(18) Liebermeister, L.; Petersen, F.; Münchow, A. v.; Burchardt, D.; Hermelbracht, J.; Tashima, T.; Schell, A. W.; Benson, O.; Meinhardt, T.; Krueger, A.; Stiebeiner, A.; Rauschenbeutel, A.; Weinfurter, H.; Weber, M. Tapered fiber coupling of single photons emitted by a deterministically positioned single nitrogen vacancy center. Applied Physics Letters 2014, 104, 031101.

(19) Patel, R. N.; Schroder, T.; Wan, N.; Li, L.; Mouradian, S. L.; Chen, E. H.; Englund, D. R. Efficient photon coupling from a diamond nitrogen vacancy center by integration with silica fiber. Light Sci Appl 2016, 5, e16032–.

(20) Politi, A.; Cryan, M. J.; Rarity, J. G.; Yu, S.; O'Brien, J. L. Silica-on-Silicon Waveguide Quantum Circuits. Science 2008, 320, 646–649.

(21) Lodahl, P.; Mahmoodian, S.; Stobbe, S. Interfacing single photons and single quantum dots with photonic nanostructures. Rev. Mod. Phys. 2015, 87, 347–400.

(22) Hausmann, B. J. M.; Shields, B.; Quan, Q.; Maletinsky, P.; McCutcheon, M.; Choy, J. T.; Babinec, T. M.; Kubanek, A.; Yacoby, A.; Lukin, M. D.; Lončar, M. Integrated Diamond Networks for Quantum Nanophotonics. Nano Lett. 2012, 12, 1578–1582.

(23) Makhonin, M. N.; Dixon, J. E.; Coles, R. J.; Royall, B.; Luxmoore, I. J.; Clarke, E.; Hugues, M.; Skolnick, M. S.; Fox, A. M. Waveguide Coupled Resonance Fluorescence from On-Chip Quantum Emitter. Nano Lett. 2014, 14, 6997–7002.

(24) Türschmann, P.; Rotenberg, N.; Renger, J.; Harder, I.; Lohse, O.; Utikal, T.; Götzinger, S.; Sandoghdar,V. On-chip linear and nonlinear control of single molecules coupled to a nanoguide. arXiv preprint arXiv:1702.05923 2017,

(25) Shen, J.-T.; Fan, S. Strongly Correlated Two-Photon Transport in a One-Dimensional Waveguide Coupled to a Two-Level System. Phys. Rev. Lett. 2007, 98, 153003–.

(26) Ralph, T. C.; Söllner, I.; Mahmoodian, S.; White, A. G.; Lodahl, P. Photon Sorting, Efficient Bell Measurements, and a Deterministic Controlled-Z Gate Using a Passive Two-Level Nonlinearity. Phys. Rev. Lett. 2015, 114, 173603.

(27) Goban, A.; Hung, C.-L.; Hood, J. D.; Yu, S.-P.; Muniz, J. A.; Painter, O.; Kimble, H. J. Superradiance for Atoms Trapped along a Photonic Crystal Waveguide. Phys. Rev. Lett. 2015, 115, 063601.

(28) Arcari, M.; S?ï£¡llner, I.; Javadi, A.; Lindskov Hansen, S.; Mahmoodian, S.; Liu, J.; Thyrrestrup, H.; Lee, E.; Song, J.; Stobbe, S.; Lodahl, P. Near-Unity Coupling Efficiency of a Quantum Emitter to a Photonic Crystal Waveguide. Phys. Rev. Lett. 2014, 113, 093603–.

(29) Hwang, J.; Hinds, E. A. Dye molecules as single-photon sources and large optical nonlinearities on a chip. New Journal of Physics 2011, 13, 085009–.

(30) Kewes, G.; Schoengen,M.; Neitzke, O.; Lombardi, P.; Schönfeld, R.-S.; Mazzamuto, G.; Schell, A. W.; Probst, J.; Wolters, J.; Löchel, B.; Toninelli, C.; Benson, O. A realistic fabrication and design concept for quantum gates based on single emitters integrated in plasmonic-dielectric waveguide structures. Scientific Reports 2016, 6, 28877–.

(31) Zadeh, I. E.; Elshaari, A. W.; Jöns, K. D.; Fognini, A.; Dalacu, D.; Poole, P. J.; Reimer, M. E.; Zwiller, V. Deterministic Integration of Single Photon Sources in Silicon Based Photonic Circuits. Nano Lett. 2016, 16, 2289–2294.

(32) Polisseni, C.; Major, K.; Grandi, S.; Boissier, S.; Clark, A.; Hinds, E. Coupling Dye Molecules to a Silicon Nitride Waveguide. Australian Conference on Optical Fibre Technology. 2016; pp AT3C–1.

(33) Trebbia, J.-B.; Ruf, H.; Tamarat, P.; Lounis, B. Efficient generation of near infrared single photons from the zero-phonon line of a single molecule. Opt. Express 2009, 17, 23986–23991.

(34) Nicolet, A. A. L.; Hofmann, C.; Kol'chenko, M. A.; Kozankiewicz, B.; Orrit, M. Single Dibenzoterrylene Molecules in an Anthracene Crystal: Spectroscopy and Photophysics. ChemPhysChem 2007, 8, 1215–1220.

(35) Toninelli, C.; Early, K.; Bremi, J.; Renn, A.; Götzinger, S.; Sandoghdar,V. Near-infrared single-photons from aligned molecules in ultrathin crystallinefilms at room temperature. Opt. Express 2010, 18, 6577–6582.

(36) Toninelli, C.; Delley, Y.; Stoferle, T.; Renn, A.; Gotzinger, S.; Sandoghdar, V. A scanning microcavity for in situ control of single-molecule emission. Applied Physics Letters 2010, 97, 021107.

(37) Ovvyan, A. P.; Gruhler, N.; Ferrari, S.; Pernice, W. H. P. Cascaded Mach-Zehnder interferometer tunable filters. Journal of Optics 2016, 18, 064011.

(38) Daveau, R. S.; Balram, K. C.; Pregnolato, T.; Liu, J.; Lee, E. H.; Song, J. D.; Verma, V.; Mirin, R.; Nam, S.; Midolo, L.; Stobbe, S.; Srinivasan, K.; Lodahl, P. Efficient fiber-coupled single-photon source based on quantum dots in a photonic-crystal waveguide. arXiv:1610.08670v1 [quant-ph] 2016,

(39) Davanco, M.; Liu, J.; Sapienza, L.; Zhang, C.-Z.; Cardoso, J. V. D. M.; Verma, V.; Mirin, R.; Nam, S. W.; Liu, L.; Srinivasan, K. A heterogeneous III-V/silicon integration platform for on-chip quantum photonic circuits with single quantum dot devices. arXiv preprint arXiv:1611.07654 2016,

(40) Ding, Y.; Peucheret, C.; Ou, H.; Yvind, K. Fully etched apodized grating coupler on the SOI platform with- 0.58 dB coupling efficiency. Optics letters 2014, 39, 5348–5350.

(41) Nakhimovsky, M. J.-D., I.; Lamotte In Handbook of Low-Temperature Electronic Spectra of Polycyclic Aromatic Hydrocarbons; 40;, P. S. D., Ed.; Elsevier, 1989.

(42) Lee, H.-W.; Kim, J. Quantum teleportation and Bell's inequality using single-particle entanglement. Phys. Rev. A 2000, 63, 012305.

(43) Chang, D. E.; Gritsev, V.; Morigi, G.; Vuletic, V.; Lukin, M. D.; Demler, E. A. Crystallization of strongly interacting photons in a nonlinear optical fibre. Nat Phys 2008, 4, 884–889.

(44) Faez, S.; Türschmann, P.; Haakh, H. R.; Götzinger, S.; Sandoghdar, V. Coherent Interaction of Light and Single Molecules in a Dielectric Nanoguide. Phys. Rev. Lett. 2014, 113, 213601.

(45) Goban, A.; Hung, C.-L.; Yu, S.-P.; Hood, J.; Muniz, J.; Lee, J.; Martin, M.; McClung, A.; Choi, K.; Chang, D.; Painter, O.; Kimble, H. Atom-light interactions in photonic crystals. Nat Commun 2014, 5, 3808.

(46) Zang, X.; Yang, J.; Faggiani, R.; Gill, C.; Petrov, P. G.; Hugonin, J.-P.; Vynck, K.; Bernon, S.; Bouyer, P.; Boyer, V.; Lalanne, P. Interaction between Atoms and Slow Light: A Study in Waveguide Design. Phys. Rev. Applied 2016, 5, 024003.

# Graphical TOC Entry

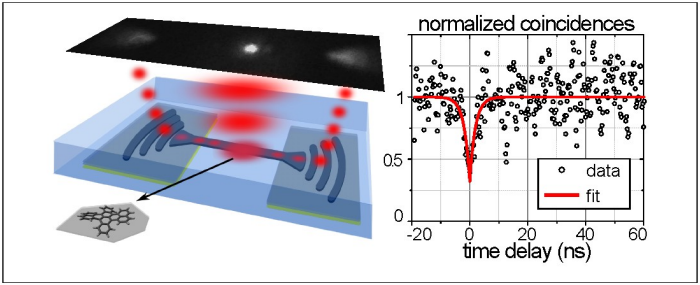